\documentclass[notitlepage,prd,superscriptaddress,nofootinbib]{revtex4-2}
\usepackage{amsmath} 
\usepackage{xcolor}
\usepackage{bbold}
\def\bz{\bar{z}}
\def\bpi{\bar{\pi}}

\def\be{\begin{equation}}
\def\ee{\end{equation}}
\def\arr{\begin{array}{rll}}
\def\ea{\end{array}}
\def\bea{\begin{eqnarray}}
\def\eea{\end{eqnarray}}

\def\N2{$N{=}2$}

\def\>{\rangle}
\def\<{\langle}
\def\+{\dagger}
\def\={\ =\ }

\begin{document}
\renewcommand{\thefootnote}{\arabic{footnote}}
\setcounter{footnote}0
\setcounter{page}{1}
\title{Note on   supersymmetric mechanics with spin-orbit interaction}
\author{Sergey Krivonos}
\affiliation{Bogoliubov Laboratory of Theoretical Physics , Joint Institute for Nuclear Research,
Dubna, Russia}
\author{Armen
Nersessian}
\affiliation{Yerevan Physics Institute,
2 Alikhanyan Brothers  St., Yerevan, 0036, Armenia}
\affiliation{Institute of Radiophysics and Electronics, Ashtarak-2, 0203, Armenia }
\affiliation{Bogoliubov Laboratory of Theoretical Physics , Joint Institute for Nuclear Research,
Dubna, Russia}
\begin{abstract}

We propose a simple model of two-dimensional $\mathcal{N}=2$ superconformal mechanics  with a spin-orbit interaction term and demonstrate that it inherits the Galilean symmetry of the initial free-particle system.
We    then propose a   quaternionic counterpart of this system, which   describes the four-dimensional  $\mathcal{N}=4$ superconformal mechanics $D(1,2|\alpha)$. The bosonic part of the Hamiltonian describes a  free particle on the cone,  while the fermionic part necessarily includes the spin-orbit interaction term.
\end{abstract}
\maketitle

%

\section{Introduction}
 Supersymmetric mechanics were initially introduced as toy models for supersymmetric field theories \cite{w78}. However, it was soon realized that these models are highly relevant to the study of numerous problems in conventional quantum mechanics (for a review, see \cite{rev} and references therein). Nevertheless, the superfield approach—inspired by supersymmetric field theories—remains the primary tool for studying of supersymmetric mechanics. Over the past two decades, significant efforts have  been  devoted to  constructing models of $\mathcal{N}\geq 4$ supersymmetric mechanics within this framework. In particular, all  $\mathcal{N}=8$ were classified in \cite{t0}. Much attention has been devoted to superconformal mechanics (see, e.g.,    \cite{FIL} and references therein).

Many of these models were formulated using Lagrangians that depend on additional bosonic variables introduced via first-order kinetic terms (see, e.g., \cite{semidyn}). These variables were interpreted by the authors as isospin (or semidynamical) variables. Recently, Fedoruk and Ivanov proposed a highly nontrivial $\mathcal{N}=8$  supersymmetric mechanical model of this kind \cite{IvFed}. The model was formulated in terms of a Lagrangian depending on a single real bosonic variable, eight complex bosonic variables, and eight real fermionic variables. Only the real bosonic variable appears with a second-order kinetic term, while the kinetic terms for the complex bosonic variables are of first order. The authors interpreted the model as an $\mathcal{N}=8$   supersymmetric extension of a one-dimensional isospin particle.

Shortly afterward, together with Erik Khastyan, we demonstrated that the Fedoruk-Ivanov model possesses dynamical superconformal symmetry and describes an  $\mathcal{N}=8$ supersymmetric extension of a free particle on an eight-dimensional cone. Additionally, its fermionic part can be interpreted as a spin-orbit coupling term \cite{kkn}. In \cite{KN24}, this construction was extended to a broader class of $\mathcal{N}=8$  superconformal mechanics. We also argued that the Fedoruk-Ivanov model has an intrinsic connection to octonions, suggesting that similar $\mathcal{N}=2,4 $  models exist in two- and four-dimensional spaces and are associated with complex numbers and quaternions, respectively. In this note, we demonstrate that this is indeed the case.

In {\sl Section 2}, we propose and describe in detail an $\mathcal{N}=2$  supersymmetric mechanics model with two bosonic and two fermionic degrees of freedom. The bosonic part of the Hamiltonian corresponds to a free particle on a two-dimensional plane, while the fermionic part defines a spin-orbit interaction term. Moreover, the system inherits the Galilean symmetry of the two-dimensional free particle and possesses the dynamical $su(1,1|1)$  superconformal symmetry.

In {\sl Section 3}, we present the quaternionic counterparts of the supercharges introduced in the previous section and construct, in this way, an  $\mathcal{N}=4$ supersymmetric mechanics model with four bosonic and four fermionic degrees of freedom. This system exhibits the dynamical superconformal symmetry $D(1,2|\alpha)$  and is formulated on a four-dimensional Euclidean cone, which reduces to $\mathbf{R}^4$  for $\alpha=1$. The fermionic part of the Hamiltonian necessarily includes a spin-orbit coupling term.

\section{$\mathcal{N}=2$ supersymmetric systems}
 Consider the two-dimensional Euclidean plane ${\mathbf{{R}}^2} = {\mathbf{C}}^1$
equipped with the metrics
\be
(ds)^2=\sum_{\mu=0,1}dx^\mu dx^\mu=2dzd\bar z,\qquad {\rm with }\quad z =\frac{x^0+\imath x^1}{\sqrt{2}}.
\ee
Then extend this space by   two fermionic degrees of freedom   $\theta^{1},\theta^2$  and introduce the momenta $p_0,p_1$ canonically conjugated with  $x^0,x^1$.
This extended superspace can be equipped with the canonical (super)symplectic structure
\be
\Omega = \sum_{\mu=0,1}\left( dp_\mu\wedge d{x}^\mu+ \frac12 d\theta^\mu\wedge d\theta^\mu\right)=
d\pi\wedge dz+d\bar{\pi}\wedge d\bar{z}+d\eta\wedge d\bar\eta,
\label{ss2}\ee
where
\be
 z=\frac{x^0+\imath x^1}{\sqrt{2}},\qquad \pi=\frac{p_0 -\imath p_1}{\sqrt{2}},\qquad  \eta=\frac{\theta^0 +\imath \theta^1}{\sqrt{2}}.
\ee
 The respective Poisson brackets are given by the non-zero  relations
 \be
 \{\pi,z\}=1,\quad \{\eta,\bar\eta\}= 1\quad \Leftrightarrow\quad  \{p_\mu,x^\nu\}=\delta_{\mu}^\nu ,\quad  \{\theta^\mu,\theta^\nu\}=\delta^{\mu\nu}.
 \ee

 Our goal is to construct, on this space, the $\mathcal{N}=2$ supersymmetric extension of the free particle system defined by the Hamiltonian  $H=\pi\bar\pi$.

For this purpose we
 choose  the following Ansatz  for supercharge:
  \be
 Q= {\rm e}^{\imath \nu\; arg\; z} \pi \eta,   \qquad {\rm with}\quad {\rm e}^{\imath \; arg\; z}=\frac{z}{|z|},\quad \nu\in \mathcal{N}.
 \label{Q2} \ee
 It immediately yields the $\mathcal{N}=2$ supersymmetric Hamiltonian
 \be
 \mathcal{H}= \pi\bpi + \frac{\nu}{2}\frac{ (\pi z -\bpi\bz) \eta\bar\eta  }{ |z|^2}\;: \qquad \{Q, \overline{Q}\}=   \mathcal{H},\quad\{Q, Q\}=0.
\ee
The Hamiltonian and supercharges  commute  with the following constants of motion:
 \bea
& {  {\Pi}}=\pi -\frac{\nu}{2z} \eta\bar\eta \;,\quad
 \mathcal{J}=  \imath(\pi z -\bar\pi\bar z) +\imath (1- \nu) \eta\bar\eta :&\\
 & \{  {\Pi},\mathcal{H}\}= \{ \mathcal{J}, \mathcal{H}\}=0,\quad   \{  {\Pi},Q\}=  \{  {\Pi},\bar Q\}=\{ \mathcal{J}, Q\} =0&
 \eea
The algebra of these constants of motion is of the form
\be
   \{ {\Pi}, {\bar\Pi}\}\approx 0,\quad\{ {\Pi}, \mathcal{J}\}=\imath\Pi,
 \label{7}\ee
 i.e.  the constructed system    inherits symmetries    of the two-dimensional free particle\footnote{The  notation $\approx$ in \eqref{7} is not a misprint
because to accurately calculate   the Poisson brackets  between $\Pi, \bar\Pi$ we should have in mind  the relation $
\frac{\partial}{\partial  z}\frac{1}{\bar z}=2\pi\imath$  so that  $\{\Pi,\bar \Pi\}=2\pi\nu\imath  \delta^2(z)\imath\eta\bar\eta  $ ( here $\pi$ is not a momentum but the irrational number:  $\pi=3,1415...$ ).
After quantization  the operator $\Pi$ acts on the upper/lower component of the two-component spinor
as $\partial_z \mp \imath\nu/z  $.
Since $A_z=\frac{\imath}{2\mathcal{\pi} z}$ is  the vector potential of the infinitely thin magnetic solenoid (in two-dimensional terms - magnetic vortex),
the system can be interpreted as an anyon  with spin  $\nu$, or, equivalently, as a charged particle moving in the field of an infinitely thin solenoid, see, e.g. \cite{bohlin}.}.

Performing canonical transformation mixing bosonic and fermionic degrees of freedom, one could represent the system  in the form of a free particle.
Namely, introducing the new fermionic variable
\be
\tilde{\eta}={\rm e}^{\imath \nu\; arg\; z}  \eta\;,
\ee
we get
\be
\{\tilde{\eta},\bar{\tilde{\eta}}\}=1,\quad  \{\Pi, z\}=1,\quad \{\bar\Pi,\bar z\}=1,  \qquad \{\Pi, \tilde{\eta}\}=\{\Pi, \bar{\tilde{\eta}}\}=0.
\ee
 In these terms the Hamiltonian, supercharges, and rotation generator read
 \be
 \mathcal{H}=\Pi\bar\Pi,\quad  Q=\Pi\tilde\eta,\quad \overline{Q}={\bar\Pi}\bar{\tilde\eta}, \quad \mathcal{J}=\imath(\Pi z-\bar\Pi\bar z)+\imath{\tilde{\eta}}\bar{\tilde{\eta}}\;.
 \ee

The above  $\mathcal{N}=2$ supersymmetry algebra can  be immediately extended to the superconformal one by adding the generators
\bea
&K= {z\bar z} ,\quad D= z\pi+\bz \bpi =z\Pi+\bz\bar\Pi  ,\quad \tilde{\mathcal{J}}= \mathcal{J}+\imath \eta \bar\eta,  \quad S  = {\rm e}^{\imath \nu\; arg\; z} \bar z \eta =\bar z\tilde{\eta} :&\\
&\{K,\mathcal{H}\}= - D,\quad \{D,K\}= 2 K,\quad  \{D,\mathcal{H}\}=- 2 \mathcal{H},&\\
& \{D,Q\}=- Q,\quad \{D,S\}= S, \quad \{K,Q\}=- S,\quad \{S,Q\}= 0, &\\
&\{S,\bar Q\}= \frac{1}{2} \left( D + \imath \tilde{\mathcal{J}}\right),\quad  \{\mathcal{H},S\}= Q ,\quad \{S, \bar S\}=   K,\quad
 \{\tilde{\mathcal{J}}, Q\}= \imath Q . &
\eea
Quantization of  similar two-dimensional model (with additional oscillator term and broken $\mathcal{N}=2$ supersymmetry)   was given in \cite{toppan}.
 
Let us complete this section by two remarks:
\begin{itemize}
\item
 The bosonic and fermionic parts of the rotational momentum $\mathcal{J}$ (and of $\tilde{\mathcal{J}}$ ) define the constants of motion as well, but they do not commute with supercharges
 \bea
& L:=\imath (\pi z -\bar\pi\bar z),
 \quad R:= \imath \eta\bar\eta\;:&\\
 &
 \quad \{L, \mathcal{H}\}= \{R,\mathcal{H}\}=\{L,R\}=0,\quad
 \{L,Q\}=\imath (\nu-1)  Q,\quad \{R,Q\}=\imath  Q.&
 \label{15} \eea
In the three-dimensional terms $L$ corresponds  to the third component of the angular momentum  while $R$ upon quantization results
in the Pauli matrix $\sigma_3$. So, the fermionic part of   the Hamiltonian defines   the spin-orbit coupling term.  It is the well-known system describing
the so-called spin-Hall effect observed two decades ago \cite{Bernevig}.
The supercharge is nothing else but the Hamiltonian of the Rashba model.
To our  knowledge,  a supersymmetric features of that model  have  not been noticed  in the  extended literature on this subject.

\item
The constructed supersymmetric system preserves all symmetry properties upon the inclusion of the constant magnetic field.
This  can be done  by an  appropriate deformation of the  symplectic structure
\be
\Omega_B=\Omega+\imath B dz \wedge d\bar z
\ee
 and of the respective Poisson brackets
\be
  \{\pi, z\}_B=1,\quad  \{\eta,\bar\eta\}_B=1,\quad \{\pi,\bar\pi\}_B= \imath B,
\ee
  where $B=const$ is the strength  of the magnetic field.

Then, choosing the supercharge in the form \eqref{Q2}, we arrive at the Hamiltonian
\be
\mathcal{H}_B:=\{Q,\overline{Q}\}_B=\mathcal{H} + \imath B  \eta\bar\eta.
\ee
In this case, the translation and rotation generators  take the  form
\be
\Pi_B=\Pi + \imath B \bar z, \qquad \mathcal{J}_B=\mathcal{J} - {B z\bar z} .
\ee
They remain commuting  with the Hamiltonian and supercharges and form  the two-dimensional  Galilieo algebra.
\end{itemize}

Let us   rewrite  the  above expressions in the polar coordinates
\be
z=\frac{r{\rm e}^{\imath\nu\varphi}}{\sqrt{2}},\quad \pi=\frac{{\rm e}^{-\imath\nu\varphi}}{\sqrt{2}} \left(p_r-\frac{\imath}{\nu}\frac{p_\varphi}{ r} \right)\;:\quad \{p_r,r\}=1, \quad\{p_\varphi,\varphi\}=1.
\ee
In these terms the Hamiltonian and supercharges take the following form
\be
\mathcal{H}=\frac{p^2_r}{2}+\frac{p^2_\varphi}{2\nu^2 r^2} +  \frac{p_\varphi \theta^0\theta^1}{2 r^2},
\qquad  Q=\frac{1}{ \sqrt{2}}\left(Q_0+\imath Q_1\right),  \qquad Q_0=p_r\theta^0+\frac{ p_\varphi \theta^1}{\nu r}, \quad Q_2=p_r\theta^1-\frac{ p_\varphi \theta^0}{\nu r}\;.
\label{qpolar}\ee
The generators of the Galilieo algebra read
\be
\mathcal{J}=\frac{p_\varphi}{\nu}+(1-\nu)\theta^0\theta^1,\quad \Pi=\Pi_0+\imath \Pi_1 =\frac{{\rm e}^{-\imath\nu\varphi}}{\sqrt{2}}
\left(p_r-\imath\frac{p_\varphi -\nu\theta^0\theta^1}{r} \right)
\ee
 These expressions will be used  in the next section to  choose the Ansatz   for the $\mathcal{N}=4$ supersymmetric mechanics model.

 \section{$\mathcal{N}=4$ supersymmetric system}
 It this Section we    propose the quaternionic counterpart  of the  $\mathcal{N}=2$ supersymmetric systems constructed in the previous Section. We expect to obtain    in this way the
 $\mathcal{N}=4$ supersymmetric extensions of the free particle systems on the four-dimensional Euclidean space
 $\mathbf{{R}}^{4}=\mathbf{{H}}^1$  equipped with the metrics
\be
(ds)^2=\sum_{\mu=0}^3dx^\mu dx^\mu =2d\mathbf{z}d\bar{\mathbf{z}},\qquad   z =\frac{x^0+\sum_{i=1}^3 \mathbf{e}_i x^i}{\sqrt{2}},
\ee
where
\be
\mathbf{e}_i \mathbf{e}_j=-\delta_{ij}+\epsilon_{ijk}\mathbf{e}_k,
\ee
and   $\epsilon_{ijk}$ is the  completely antisymmetric tensor with  $\epsilon_{123}=1$.

Then we extend this  space  by four fermionic   degrees of freedom defined by the coordinates $\theta^{0},\theta^i$ and introduce the momenta $p_0,p_i$ canonically conjugated with  $x^0,x^i$.

This extended superspace can be equipped with  the (super)canonical
Poisson brackets   given by the following nonzero  relations:
 \be
 \{p_\mu,x^\nu\}=\delta_{\mu}^\nu ,\qquad  \{\theta^\mu,\theta^\nu\}=\delta^{\mu\nu}.
 \ee
 The respective (super)symplectic structure can be re-written in quaternionic coordinates in a form similar to \eqref{ss2}
   \be
\Omega = \mathbf{Re}\;\left(d\boldsymbol{\pi} \wedge d\mathbf{z} + d\boldsymbol{\eta}\wedge d\boldsymbol{\bar\eta}\right)=\sum_{\mu=0}^{3}\left( dp_\mu\wedge d{x}^\mu +\frac12 d\theta^\mu \wedge d\theta^\mu\right) ,
\label{ss4}\ee
where
\be
\boldsymbol{\pi}=\frac{ p_0-\sum_{i=1}^{3}p_i\mathbf{e}_i }{ \sqrt{2}}, \quad \boldsymbol{z}=\frac{ x^0+\sum_{i=1}^{3}x^i\mathbf{e}_i}{\sqrt{2}},\quad
\boldsymbol{\eta}=\frac{ \theta^0+\sum_{i=1}^{3}\theta^i\mathbf{e}_i}{\sqrt{2}}.
\ee
Before extending the Anzats \eqref{Q2} to the quaternionic case, we
present   some  quaternionic relations
 that  will be further used for the construction of $\mathcal{N}=4$ supersymmetric mechanics.

At first, let us write down the relations on bosonic  variables
\be
\frac{\boldsymbol{z\pi}}{|\mathbf{z}|}=\frac{1}{\sqrt{2}}\left(p_r+\frac{L^+_i\mathbf{e}_i}{r}\right), \qquad  \frac{\boldsymbol{     \pi z}}{|\mathbf{z}|}=\frac{1}{\sqrt{2}}\left(p_r+\frac{L^-_i\mathbf{e}_i}{r}\right),
\ee
 where
 \be r:=\sqrt{2}|\mathbf{z}|=\sqrt{\sum_{\mu=0}^3 x^\mu x^\mu},\quad p_r=\sqrt{2}\frac{\mathbf{Re}\; \boldsymbol{ z\pi}}{|z|}=\frac{ \sum_{\mu=0}^3 p_\mu x^\mu}{r}, \label{pr} \ee
 and
\be
L^{\pm}_i=p_0x^i-p_ix^0\pm \epsilon_{ijk}p_j x^k\;: \{L^{\pm}_i, L^{\pm}_j\}=\pm 2  \epsilon_{ijk}L^{\pm}_k, \quad \{L^+_i,L^-_j\}=0.
\label{lpm}\ee
So $L^\pm_i$ are  the  generators of the $so(3)$  algebras corresponding to the decomposition $so(4)=so(3)\times so(3)$ with the
  Casimirs
\be
 \mathcal{C}_{so(3)}=\sum_{i=1}^3 (L^{\pm}_i)^2 = \left(\sum_{\mu =0}^3 p^2_\mu \right)  \left(\sum_{\mu=0}^3 (x^\mu)^2 \right) - \left(\sum_{\mu=0}^3 p_\mu x^\mu  \right)^2.
\label{casimir}\ee
Hence  $\mathcal{C}_{so(4)}=2\mathcal{C}_{so(3)}$.

 The Poisson brackets between  the functions \eqref{pr}  and the $so(3)$ generators \eqref{lpm} are as follows:
\be
 \{p_r, r\}=1,\qquad \{L^\pm_i,r\}=\{L^\pm_i,p_r\}=0. 
\ee

 %

Let also introduce the fermionic functions
\be
\mathbf{R}^+=    \boldsymbol{\bar\eta}\boldsymbol{\eta}=R^+_{i}\mathbf{e}_i,\quad    \mathbf{R}^-= \boldsymbol{\eta}\boldsymbol{\bar \eta} =R^-_{i}\mathbf{e}_i,\quad \mathbf{R}^0=\boldsymbol{\eta}^2=\boldsymbol{\bar\eta}^2= \frac12 R^0_i\mathbf{e}_i\ee
with
\be
R^\pm_{i}= \theta^0\theta^i\mp \frac12\epsilon _{ijk}\theta^j\theta^k , \quad R^0_{i}=  R^-_i -R^+_i  =\epsilon_{ijk}\theta^j\theta^k.
\ee
They obey the following Poisson bracket relations:
\bea
&\{R^\pm_i, R^\pm _j\}=\pm 2\epsilon_{ijk}R^\pm_k,\quad \{R^+_i, R^-_j\}=0,\quad 
\{R^\pm_i, R^0_j\}=-2 \epsilon_{ijk}R_k^{\pm},\quad \{R^0_i, R^0_j\}=-2\epsilon_{ijk}R_k^{0}&
 \eea
i.e., generate $so(4)$ (and $so(3)$) transformations in the fermionic sector.

It is instructive to write down the component expressions of the three-fermionic    functions
\be
\boldsymbol{\Lambda}= \boldsymbol{\eta\bar\eta\eta}=\frac{\Lambda_0+\mathbf{e}_i\Lambda_i}{6\sqrt{2}}\; :\; \Lambda_0= \frac13\epsilon_{ijk}\theta^i\theta^j\theta^k,\quad \Lambda_i=
- \epsilon_{ijk}\theta^j\theta^k\theta^0,\qquad
\{\Lambda_0,\Lambda_i\}=\{\Lambda_i,\Lambda_j\}=0
\ee

With these expressions at hand we are ready to construct the $\mathcal{N}=4$ supersymmetric mechanics on $\mathbb{R}^4=\mathbb{H}^1$.\\

 Due to noncommutativity of quaternions,  there are many ways to extend the Ansatz \eqref{Q2} to the quaternionic case, including its extension by   terms trilinear on  fermions ,
   e.g.,
 \be
 \boldsymbol{Q}= \left(\frac{\mathbf{z}}{|\mathbf{z}|}\right)^{\nu_1} \boldsymbol{\pi}\left(\frac{\mathbf{z}}{|\mathbf{z}|}\right)^{\nu_2} \boldsymbol{\eta}  +\mu \left(\frac{\mathbf{z}}{|\mathbf{z}|}\right)^{\nu_3}\frac{\boldsymbol{\Lambda}}{|\mathbf{z}|} .
\quad\ee
However, the absence of differentiation by quaternionic  variables makes them not very effective for our purposes.

The simplest  quaternionic analog of  the     $\mathcal{N}=2$ supercharge with $\nu=1$ \eqref{Q2} reads:
\be
\mathbf{\mathcal{Q}}= \frac{\mathbf{z}}{|\mathbf{z}|} \boldsymbol{\pi}\boldsymbol{\eta}=\frac{1}{\sqrt{2}}\left(\mathcal{Q}_0+\mathcal{Q}_i\mathbf{e}_i\right)\;:\qquad
 \mathcal{Q}_0  =  p_r \theta^0 +\frac{L^+_k \theta^k}{ r},\quad
\mathcal{Q}_i = p_r \theta^i - \frac{L^+_i \theta^0 -\epsilon_{ijk} L^+_j \theta^k}{ r}
\ee
However,one can check that these functions do not form  the $\mathcal{N}=4$ superalgebra.

To get the  $\mathcal{N}=4$ superalgebra, we choose a more general Ansatz  including the three-fermionic term and depending on the coefficient  $\alpha $ which could be viewed as the  analog of the coefficient $\nu$ in \eqref{Q2}(cf. \eqref{qpolar}).

Then,  after direct calculation we get the expressions for supercharges which obeys $\mathcal{N}=4$ supersymmetry algebra
\be
Q_0  =  p_r \theta^0 - \alpha\frac{   L^+_k \theta^k  }{r}+ (1+2\alpha)\frac{\Lambda_0}{2r}  ,\qquad
Q_i = p_r \theta^i +\alpha \frac{  L^+_i\theta^0 - \epsilon_{ijk} L^+_j \theta^k}{r}
+(1+2\alpha)\frac{\Lambda_i}{2r}  ,
\label{N40}\ee
Namely,  
\be
\{Q_0, Q_0\}=2\mathcal{H}, \quad \{Q_i, Q_j\}=2\delta_{ij}\mathcal{H},\qquad \{Q_0, Q_i\}=0,
\label{n4}\ee
where the Hamiltonian is given by the expression
\be
\mathcal{H}=\frac{1}{2} p_r^2 +\frac{1}{2r^2}\left[{\alpha^2}  { L^+_k L^+_k}  + 2\alpha { L^+_k R^+_k}
-(1+2\alpha) \Gamma \right],\qquad {\rm with}\quad
 \Gamma :=\mp \frac13  {R^{\pm}_k R^{\pm}_k}= \frac13 {\theta^0\epsilon_{ijk}\theta^i\theta^j\theta^k} .
\label{N42} \ee
%
It is an obvious that
the following sets of $so(3)$ generators define the constants of motion of the Hamiltonian
\be
L^-_i,\quad R^-_i,\quad \mathcal{J}^+_i=L^+_i+ R^+_i\;:\quad\{ {L}^-_i, \mathcal{H}\}=
\{R^-_i, \mathcal{H}\}=\{\mathcal{J}^+_i, \mathcal{H}\}=0.
\ee
Their
 Poisson brackets with supercharges are as follows:
\bea
&&\{\mathcal{J}^+_i,Q_0\}=-Q_i,\quad  \{ \mathcal{J}^+_i,Q_j,\}=\delta_{ij}Q_0+\epsilon_{ijk}Q_k,\\
&&\{{R}^-_i,Q_0\}=-Q_i,\quad
 \{{R}^-_i,Q_j\}=\delta_{ij}Q_0-\epsilon_{ijk}Q_k,\\
&& \{Q_0, L^-_k\}= \{Q_i, L^-_k\}=0,
\eea
This  superalgebra can be expended to the superconformal algebra $D(1,2|\alpha)$
by introducing the generators  of dynamical superconformal symmetry
\be
K=\frac{r^2}{2},\quad
 D= \frac{r p_r}{2},  \quad
S_0 = r \theta_0, \quad S_i = r \theta_i.
\ee
Their Poisson brackets with the former generators are as follows:
\bea
& \left\{S_0,Q_0\right\} = 2 D, \quad \left\{S_0,Q_i\right\} =\alpha {\cal J}^+_i -(1+\alpha) R^-_i,&
 \\
& 
\left\{S_i,Q_0\right\} = -\alpha {\cal J}^+_i +(1+\alpha) R^-_i, \quad
\left\{S_i,Q_j\right\} = 2 \delta_{ij} D - \epsilon_{ijk}( \alpha {\cal J}^+_k +(1+\alpha)R^-_k), &\\
&   \left\{\mathcal{J}^+_i,S_0\right\} =-S_i, \quad \left\{\mathcal{J}^+_i,S_j\right\} =\delta_{ij}S_0+ \epsilon_{ijk}S_k,\quad 
\left\{R^-_i,S_0\right\} =-S_i,
\quad
\left\{R^-_i,S_j\right\} =\delta_{ij}S_0 - \epsilon_{ijk}S_k.&
\eea

The bosonic part of the Hamiltonian in  the initial Cartesian coordinates reads
\be
H_0:= \frac{1}{2} p_r^2 +\frac{ {\alpha^2}  L^+_k L^+_k }{2r^2}=   \frac{\alpha^2}{2} \left(\sum_{\mu=0}^3 p_\mu p_\mu-\frac{1-\alpha^2}{\alpha^2}\sum^3_{\mu=0}\left( \frac{ p_\mu x^\mu}{|x|}\right)^2\right).
\ee
So  it describes the particle on the four-dimensional space equipped with the metrics
\be
(ds)^2 =  \sum_{\mu=0}^3dx^\mu dx^\mu -(1-\alpha^2)\sum_{\mu=0}^3\left(\frac{x^\mu dx^\mu}{|x|}\right)^2=\frac{\sum_{\mu=0}^3dy^\mu dy^\mu}{|y|^{2(1-1/|\alpha|)}},
\ee
where $y^\mu=|x|^{|\alpha|-1}x^\mu$.
\\

Let us complete this section by the following remarks
\begin{itemize}

\item For $\alpha=- 1$, i.e. for the $su(1,1|2))$ superalgebra,  we get the system on the four-dimensional Euclidian space, while for $\alpha^2\neq 1$ we deal with the  free particle on four-dimensional cone.

\item  For $\alpha=-1/2$, i.e., for the $osp(4|2)$ superalgebra,   the four-fermionic (i.e spin-spin interaction) term vanishes and  supersymmetry is provided solely by spin-orbit interaction.

\item The suggested model is a  a particular example of $\mathcal{N}=4$ supersymmetric mechanics  with the spin-orbit interaction term. For example,  one can choose a more general Ansatz  for supercharges, including both sets of $so(3)$ generators: not only $L^+_\mu, R^+_\mu$ but $L^-_\mu, R^-_\mu$ as well. We are planning to  consider such models elsewhere.

\item We expect that a similar model could be constructed  on the curved $so(4)$-invariant spaces by   modifying the expressions for superchages and Hamiltonian presented in \cite{bny}.
     Clearly, in this  case we will loose the dynamical superconformal symmetry.

\end{itemize}
 
\acknowledgements 
The work of A.N. was partially  supported  by the
 Armenian State Committee of Higher Education and Science within the project  21AG-1C062 .
\\

{\bf \large{Note added.}}
 After the publication of this article in {\sl Physics of Particles and Nuclei Letters}, we noticed that the model given by expressions \eqref{N40}–\eqref{N42} represents a particular realization of the $D(1,2|\alpha)$ superconformal mechanics proposed by Anton Galajinsky in \cite{Gala }.

\end{document}